\documentclass[twocolumn,showpacs,preprintnumbers,amsmath,amssymb]{revtex4}
\usepackage{graphicx}% Include figure files
\usepackage{dcolumn}% Align table columns on decimal point
\usepackage{bm}% bold math

\begin{document}

\title{THz amplification  via a Raman scattering}% \\

\author{S. Son}
\affiliation{169 Snowden Lane, Princeton, NJ, 08540}
%\author{Sung Joon Moon}
%\affiliation{PACM, Princeton University, Princeton, NJ 08544}
%\author{J.~Y. Park}
%\affiliation{Los Alamos National Laboratory}

\begin{abstract}
A THz light amplifier and the corresponding methods are proposed, based on the Raman interaction between a THz light and  visible light lasers. 
Two lasers, with their frequencies differing by that of a  Langmuir wave,  counter-propagate inside a background plasma, exciting a rather strong Langmuir wave.    
The THz light amplification occurs  through 
the non-resonant Raman scattering between the THz light  and one of the lasers in the presence of the mentioned Langmuir wave. 
It is normally the case that the non-resonant scattering does not exchange the energy between E \&M fields  but the author demonstrates that it is  possible  under the pre-exciting Langmuir wave.
%  to flow the energy from the laser to the THz light. 
%even througth    the non-resonant scattering 
% by the laser and the THz light 
%can exchange the energy  under the pre-exciting Langmuir wave.
%The key idea is that 
%can be made to have a favorable phase under the pre-exciting Langmuir wave, for the benefit of the THg light amplification (visible light laser decay).
%The amplification occurs  through the non-resonant Raman scattering between the THz light  and one of the lasers;  
%In this 
%The situation that the author considers is  %via the backward Raman scattering is proposed.  
In the presence of a  pre-existing Langmuir wave,
the ponderomotive density perturbation between the THz light and the visible-light laser is induced with the phase suitable for transferring the energy from the visible light laser to the THz light.   
%the desirable phase of the ponderomotive density perturbation under a certain condition. % for the benefit of the THz light amplification (visible light laser decay).
 %It is proposed that the density response from the ponderomotive interaction between the THz light and the laser can be  engineered under the pre-exciting Langmuir wave for the benefit of the THg light amplification (visible light laser decay). 
The condition for achieving the favorable phase  and the amplification strength, when the condition is met, is estimated as an mathematical formula.  
%The density response from the ponderomotive interaction between the THz light and the laser is engineered so that 
%it has a desired phase  to amplify (decay) the THz light (visible light laser), 
%which  is made possible under the  pre-exciting Langmuir wave. 
%The beating of the Langmuir wave and the density perturbation from the ponderomotive potential could lead to the desriable phase-lock.  
The gain is as high as or even higher than  100 per centimeter.

%A TH%z light amplification is proposed by the Raman interaction between a THz light and a visible light laser.  The desirable phase-lock of the density perturbation between a THz light and a laser could be created by a pre-exciting Langmuir wave by two lasers. 
%The beating of the Langmuir wave and the density perturbation from the ponderomotive potential could lead to the desriable phase-lock.  
%The gain could be as high as 100 per centimeter. 
%A new scheme for soft x-ray lasers  is proposed. The backward Haman scattering between an intense visible-light laser and a relativistic electron beam results in soft x-ray light via the Doppler shift. One of the most intense soft x-ray light sources is contemplated.  
\end{abstract}


\maketitle

\section{Introduction}

For a few decades,
there have been  growing 
 interests for a  commercially-viable  THz light source, as it would be critical in the fusion plasma diagnostic, the molecular spectroscopy, the tele-communication and many others~\cite{siegel,siegel2,  siegel3, booske,radar, diagnostic, security}. 
Numerous THz light sources have been invented~\cite{Tilborg,Zheng,Reimann, gyrotron, gyrotron2, gyrotron3,magnetron, qlaser, qlaser3, freelaser, freelaser2, colson, songamma,Gallardo,sonttera} and 
% including the free electron laser, the gyrotron, the vacuum electronic device, the quantum cascade laser and the laser-based technology. 
in particular, the great advance has been made in the laser-based technology~[2].
However, 
THz light source 
is not still intense  enough for many applications and  
the current inability to produce intense THz light   %to the theoretical limit 
is referred to as the ``THz Gap''~[1].   
%This is a major hurdle in commercialization.  
%It is essential to increase the intensity of the THz light before any meaningful progresses in commercialization. 
%\patentParagraph Another serious  challenge  is the high operating (initial construction) cost;
% increasing the low conversion efficiency from the input energy into the THz light. 
In addition, current light sources often require  expensive strong magnets and accelerators, or often need to be operated in extremely low-temperature. 
%The conversion efficiency from the input energy into the THz light,  currently as low as 0.01 percents, renders the operating cost even higher.
%%can be reduced further by building a compact and mobile THz light source or increasing  the conversion efficiency from the input energy into the THz light, which  is  
% as low as 0.01 percents.
%  makes the operating cost even higher. 
Significant progresses, in enhancing the intensity (power), reducing the cost (size) or increasing the efficiency,  are necessary for commercial applications of the THz light. 
If a comparison is made between the progress in the visible light laser and the THz light laser,  one obvious missing ingredient in the THz light is an amplifier, wherein a small signal gets amplified to an  intense one. 
%While quantum cascade laser is an amplifier, it only works in a low-intensity laser. 
If an intense amplifier is available, many obstacles for the THz applications could be overcome. One such amplifier is proposed in this paper based on the Raman scattering.

In the backward Raman scattering (BRS) between two lasers, %the energy is channeled from higher frequency laser to the lower frequency laser. 
the ponderomotive force between these lasers excite a Langmuir wave, which is phase-locked  to the laser ponderomotive interaction  
by a quarter cycle. 
Due to this phase lock, 
the beating current of this Langmuir density and the  laser quiver transfers the energy from the higher frequency laser to the lower frequency laser.  %the other laser.
It is a tempting idea that the Raman scattering could 
rotate the energy from the visible light laser to the THz light in a similar fashion. 
%to use the Raman scattering to amplify the THz light, extracting the energy from the visible light laser. 
%This is not easy because the resonant interaction is not achievable due to the big difference in the frequency between the THz light and the visible light laser. 
 %If they do not satisfy the resonance condition, 
%the phase of the  excited density is in sync with the ponderomotive force
%and there is no energy transfer. 
%It is a tempting  idea to use the Raman scattering to amplify the THz light, extracting the energy from the visible light laser. 
When the THz light interacts with the visible light laser, however, 
their beating ponderomotive force cannot be  
resonant,  due to their big  frequency difference.  If it is not  resonant,
  the phase of the  excited density is in sync with the ponderomotive force, resulting  no energy transfer between the laser and the THz light.  Because of this consideration, the non-resonant Raman scattering between a laser and THz light is  not seemingly suitable for the THz light amplification.

The main idea  of this paper is to circumvent this problem
utilizing the pre-existing background Langmuir wave;   
%for the Raman scattering
% transfering  the energy from the laser into the THz light. 
%Unfortunately, as explained,  the straightforward Raman scattering between a laser and THz light is not a feasible way to transfer energy because 
%\patentParagraph
%The major reason that the Raman scattering between a laser and THz light cannot be used to amplify the THz light is that 
%their beating pondermotive interaction cannot interact with the plasma  
%resonantly,  due to their big  frequency difference.
%In this paper, one method for the phase-locking is introduced. 
 Two-counter propagating lasers o excite a Langmuir wave and  
the secondary interaction of this Langmuir wave and ponderomotive force between the lasers and the THz light creates a density perturbation with the desirable phase. 
 The condition for achieving the mentioned favorable phase  and the amplification strength, when the condition is met, is estimated.
%It is for the first time that such an amplifying mechanism is identified. 
The amplification strength is very strong, reaching the gain-per-length as high as or even higher than 100 per centimeter.

\section{ Raman scattering and the energy transfer between the lasers}

Let us briefly revisit the backward Raman scattering physics. 
Consider two lasers with the frequency $\omega_1 $ and $\omega_2$ counter-propagating in the z-direction, where $\omega_1 < \omega_2$. For simplicity, let us assume that the lasers are linearly polarized and  their electric fields are parallel to each other. Their electric field is then given as 
$E_{x1} = E_1 \exp \left(\omega_1 t - k_1 z\right)$ and $E_{x2} = E_1 \exp \left(\omega_2 t + k_2 z\right)$.
Those two lasers excites the density perturbation in the z-direction via the ponderomotive interaction.   
 The density response to the ponderomotive force can be obtained from the continuity equation and the momentum equation: 

\begin{eqnarray}
 \frac{\partial \delta n_e }{\partial t}  &=&-  \mathbf{\nabla} \cdot ( \delta n_e \mathbf{\mathrm{v}} ) \nonumber \mathrm{,}\\ \nonumber \\
 m_e \frac{d \mathbf{v} }{dt} &=& e\left( \mathbf{\nabla} \phi  - \frac{\mathbf{v}}{c} \times \mathbf{B}\right) \nonumber \mathrm{,} \\ \nonumber
\end{eqnarray}
Combining the above equations with the Poisson equation $\nabla^2 \phi = -4 \pi \delta n_e e $, the density response is governed by~\cite{McKinstrie}

\begin{eqnarray} 
\left(\frac{\partial^2 }{ \partial t^2 } + \omega_{\mathrm{pe}}^2\right)\delta n_e &=&
\frac{en_0 }{m_ec} \mathbf{\nabla} \cdot \left( \mathbf{v} \times \mathbf{B} \right)\mathrm{.} \nonumber \\ \nonumber \\
&=&  -n_0 (ck_1 + ck_2)^2 a_1^* a_2\mathrm{,}
 \label{eq:den}
\end{eqnarray} 
where $\omega_{\mathrm{pe}}^2= 4\pi n_0 e^2/ m_e$ is the plasma Langmuir wave frequency, $n_0$ is the background electron density,  all physical quantities is expressed as $b(z,t) = b \exp\left(i \omega t - k z\right) + b^*\exp\left(-i \omega t + k z\right) $, $a_{1,2} = eE_i/m\omega_{1,2} c $ is the laser quiver velocity normalized by the velocity of the light and 
$E_{i}$ is the electric field of the laser $i$.

The density perturbation from the ponderomotive force of  any one pair of the lasers ($1,2$) can be  estimated from Eq.~(\ref{eq:den}). For an example,  the density perturbation from $1,2$ with the frequency $\omega_{\mathrm{pe}} \neq  \omega_2 - \omega_1 $ is given

\begin{eqnarray} 
\delta n(\omega_2 - \omega_1, k_1 + k_2)  &=& -n_0 \frac{(ck_1 + ck_2)^2 }{ (\omega_1-\omega_2)^2 - \omega_{\mathrm{pe}}^2} a_1^* a_2  \nonumber \\ \nonumber \\
&=& -n_0 C_B a_1^* a_2 \mathrm{,} \label{eq:nonresonance}
\end{eqnarray} 
where  $C_B = (ck_2 + ck_1)^2 /( (\omega_1-\omega_2)^2 - \omega_{\mathrm{pe}}^2) \gg 1$ is a positive real constant, because $\omega_{1,2} \gg \omega_{\mathrm{pe}}$.  
%Similarly,  the density perturbation for $\omega_n = \omega_1-\omega_2\neq \omega_{\mathrm{pe}}$ is given as 
%\begin{equation}
% \delta n(\omega_1 + \omega_2, k_1 + k_2) (= -n_0 (\omega_n^2/ (\omega_n^2 -\omega_{\mathrm{pe}}^2)) a_1 a_2^* \mathrm{,} \label{eq:nonresonance}
%\end{equation} 
%where  $a_2^{*}$ is the complex conjugate.  
The lasers will respond to the density perturbation in Eq.(\ref{eq:den}) by~\cite{McKinstrie}: 
%Using Eq.~(\ref{eq:den2}) and   the backward Raman scattering theory, the non-linear convection equation can be derived by taking out the fast time scale~\cite{McKinstrie};  

\begin{eqnarray}
L_1 a_1  &=& +i \frac{\omega_{\mathrm{pe}}^2 }{2\omega_1} \left( \frac{\delta n^*}{n_0} a_2\right)    \label{eq:1} \mathrm{,}\\ \nonumber \\ 
L_2 a_2  &=& +i \frac{\omega_{\mathrm{pe}}^2 }{2\omega_2} \left( \frac{\delta n}{n_0} a_1\right)    \label{eq:2} \mathrm{,}  
\end{eqnarray}
where $L_{i} = (\partial/\partial t) + v_{i} (\partial/\partial z) $ with $v_{i}$ is the group velocity of the laser. 
In case  $ \omega_2-\omega_1\neq \omega_{\mathrm{pe}}$, the density perturbation is  $\delta n \cong -C_Ba_1^* a_2 n_0$. Then,  
    $L_1 a_1  =  i(\omega_{\mathrm{pe}}^2 / 2 \omega_1) C_B|a_2|^2 a_1$; 
the energy $|a_1|^2$ does not change and only the phase of the laser is modulated. 

However, in the case $ \omega_2-\omega_1 = \omega_{\mathrm{pe}}$, 
the resonance occurs and the density perturbation is given as~\cite{McKinstrie}

\begin{equation}
L_1 \delta n  = +\mathbf{i} \frac{(c k_1 + c k_2)^2 }{2\omega_{\mathrm{pe}}}  a_1^* a_2    \label{eq:resonance} \mathrm{,}
\end{equation}
Then, $\delta n\cong i C_B(t) a_1^* a_2$, where $C_B(t)>0$ is the real number. This density perturbation leads from Eq.~(\ref{eq:2})
 $L_1 a_1  =   C_B(t)(\omega_{\mathrm{pe}}^2 / 2 \omega_2)|a_2|^2 a_1$, where $C_B>0$ is the real number. 
The energy of the laser $\omega_1$ grows while the energy of the laser $\omega_2$ decays. 
So, this density response lagging by a quarter cycle (or imaginary response)  is key for channeling energy from one laser to the other.  
The process for this out-of-phase response can be easily explained by a harmonic oscillator. Consider a simple harmonic oscillator; 

\begin{equation}
\left(\frac{d^2  }{d t^2} +\omega_0^2\right) a  = A \cos(\omega t) \mathrm{.}
\end{equation}
If $\omega_0 \neq \omega$ ($\omega_0 = \omega$), then the particular solution is $a = A/(\omega_0^2 -\omega_1^2) \cos(\omega t) $ ( $a = (A/\omega) t \sin(\omega t) $). This phase change from the cosine function to the sine function is  what it needs to happen for the energy channeling between the lasers. 

It would be desirable if a energy channeling between the THz light and one laser is possible by the above physics of Raman scattering.  However, in the plasma where the THz light can propagate, $\omega_1 \gg \omega_{\mathrm{pe}} $ and  
$\omega_1 - \omega_3 \gg \omega_{\mathrm{pe}} $, wherein  $\omega_3$ is the frequency of the THz light.  
%Under a normal circumstances, it is impossible to channel energy from the laser to the THz light becuase the resonant interaction is not feasible. 
%The current paper proposes one method to enable  the density response, even   to the non-resonant ponderomotive interaction,  
%have  a phase desirbale to the THz light amplifiction.  

\begin{figure}
\scalebox{0.3}{
\includegraphics{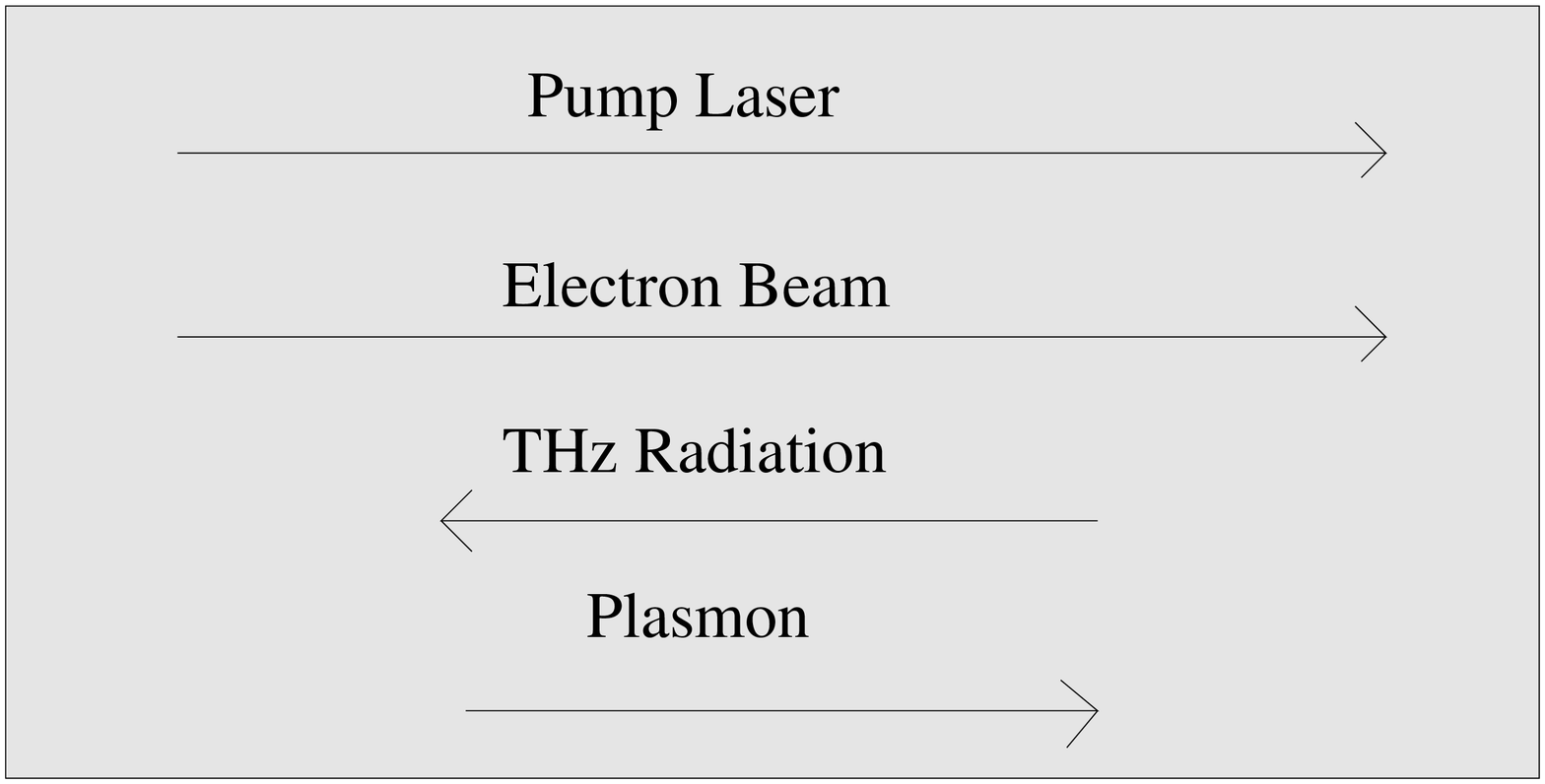}}
\caption{\label{fig:1}
The schematic diagram  about the propagation direction of the BRS pump, THz pulse, the electron beam and the plasmon. 
}
\end{figure}

\section{THz light amplification}

Consider two lasers with the frequency $\omega_1 $ and $\omega_2$ counter-propagating in the z-direction, where $\omega_2- \omega_1 =\omega_{\mathrm{pe}}$.  The density perturbation $\delta n$ will respond to the beating ponderomotive force as given in Eq.~(\ref{eq:resonance}). 
As the density excited  reaches a considerable level,  
the assumption that $\delta n \ll n_0 $ might not be valid any more, 
in which situation  the zeroth order density would be given as $n = n_0 + \delta n = n_0 + n_L(\omega_{\mathrm{pe}}, k_1+k_2) \cong n_0 +iC_B a_1^* a_2$.
%This imaginary phase of the Langmuir wave relative to the ponderomotive interaction can be utilized for the phase-lock. 
Let me assume a THz light with the frequency $\omega_3$ and the wave vector $k_3$ is moving in the same direction with $\omega_1$ laser. See Fig.~(\ref{fig:1}) for the directions of lasers and THz light. 
The density response of the plasma to the ponderomotive interaction between the THz light and two lasers are what we want to estimate. 
%With a view to how the density will respond to the ponderomotive potential between a laser and THz light,   
The expansion of the continuity equation  is given as

\begin{eqnarray}
\left(\frac{\partial^2  }{\partial t^2} + \omega_{\mathrm{pe}}^2\right)\delta n_e  &=&-  \mathbf{\nabla} \cdot ( \frac{\partial n_L}{\partial t} \mathbf{v} ) -  \mathbf{\nabla} \cdot ( (n_0+n_L)\frac{\partial \mathbf{v}}{\partial t} ) )
\nonumber \\ \nonumber \\
 &\cong&   -  (n_0 + n_L) \mathbf{\nabla} \cdot (\frac{\partial \mathbf{v}}{\partial t}) -  \left(\mathbf{\nabla} n_L\right)\cdot  (\frac{\partial \mathbf{v}}{\partial t} ) \mathrm{.} \nonumber \\ \nonumber \\ \label{eq:exp}  
\end{eqnarray}
The term involving $ (\partial n_L/\partial t) $ in Eq.(\ref{eq:exp}) is smaller than other terms by $\omega_{\mathrm{pe}}/\omega_{1,2}$, which will be ignored.  Then, there will be three remaining terms on the right side of Eq.~(\ref{eq:exp}).  
The first term involving $n_0$ is the usual term of the density response to the ponderomotive force without the presence of the Langmuir wave. 
The second term with $n_L$ and the third term with $\nabla (n_L)$ is what we will compute now.

In the above equation,
the velocity $\mathbf{v}$ is excited by the ponderomotive force between a laser and one laser. % while $n_L$ is the Langmuir wave.
There are four ponderomotive interaction ($a_3 a_1 $, $a_3 a_1^*$,  $a_3 a_2 $, $a_3 a_2^*$), which could interact with $n_L$ or $n_L^*$. Then, 
there are eight density perturbations, ($n_La_3 a_1 $, $n_La_3 a_1^*$,  $n_La_3 a_2 $, $n_La_3 a_2^*$, $n_L^*a_3 a_1 $, $n_L^*a_3 a_1^*$,  $n_L^*a_3 a_2 $, $n_L^*a_3 a_2^*$). Those eight density perturbations beat with the laser quiver ($a_1$, $a_1^*$, $a_2$, $a_2^*$). 
Therefore, there are 32 combinations. 
In the end, the author's  interest is how the THz light $a_3$ evolve and, therefore,  
the whole beating must have the total momentum $k_3$ and the frequency $\omega_3$.  There are four such combinations $(n_L [a_1 a_3] a_2^*, n_L [a_2^* a_3] a_1, 
n_L^* [a_1^* a_3] a_2, n_L [a_2 a_3] a_1^*)$, where
$n_L [a_1 a_3] a_2^*$, for an example, represents a density perturbation, by the beating of a Langmuir wave $n_L$ and the density perturbation to the ponderomotive interaction by $a_1 a_3$,  making a beat current with the laser quiver $a_2^*$.

%The situation of our interest is when the THz ligh generate a density perturbation by a ponderomotive potential with one of the laser. For the arugment, let us choose the laser $\omega_1$.  This pondermotive potential will generate a density perturbation $\delta n$ by the beating terms as given in Eq.(\ref{eq:exp}). 
%In turn, this density perturbation will generate a beat current with the laser $\omega_2$ and this beat current will have the frequency and wave vector of the original THz light, which could be either damping or amplifying or just phase modulation. As one of the beating term has the imaginary phase $n_L$, it will be either decay or amplification instead of phase modulation. 
With the above consideration in mind, by computing all four terms, 
the relevant density response for the second term in Eq.~(\ref{eq:exp}) is given as 
\begin{eqnarray}
\delta n_1 &=& + in_0\left(-A_1 |a_1|^2 a_2 + A_2 |a_2|^2 a_1^* \right) a_3 \nonumber \\   \nonumber \\ 
&=&
 +in_0\left(- A_3 |a_1|^2 a_2^* + A_4 |a_2|^2 a_1 \right)a_3 \mathrm{,} \label{eq:first}
\end{eqnarray}
where $A_1 =  (ck_1 + ck_3)^2/(\omega_2+\omega_3)^2$ represents the process of 
the ponderomotive force by $a_3 a_1 $ and producing the THz current by beating with the $a_2^*$ quiver and  $n_L$,
 $A_2 = (ck_2 + ck_3)^2/(\omega_1-\omega_3)^2$ represents
the ponderomotive force being excited by  $a_3 a_2^* $, producing 
 the THz current with the $a_1$ quiver and $n_L$, 
$A_3 = (ck_1 - ck_3)^2/(\omega_2-\omega_3)^2$ represents
the ponderomotive force by $a_3 a_1^* $, producing 
 the THz current by beating with the $a_2$ quiver and  $n_L^*$ and 
%which will produce the Thz current by beating with the $a_2$ quiver, 
 $A_4 = (ck_2 - ck_3)^2/(\omega_1+\omega_3)^2$ represents the ponderomotive force by  $a_3 a_2 $  producing the THz current by beating with the $a_1^*$ quiver and $n_L^*$.
The  relevant density response for the third term is given as 
\begin{eqnarray}
\delta n_2 &=& + in_0\left(B_1 |a_1|^2 a_2 - B_2 |a_2|^2 a_1^* \right)  \nonumber \\   \nonumber \\ 
&=&
 +in_0\left( B_3 |a_1|^2 a_2^* - B_4 |a_2|^2 a_1 \right)a_3 \mathrm{,} \label{eq:second}
\end{eqnarray}
where $B_1 = C (ck_1 + ck_3)(ck_1+ck_2)/(\omega_2+\omega_3)^2$ represents
the ponderomotive force by $a_3 a_1 $ producing the THz current by beating with the $a_2^*$ quiver and $n_L$,
%which will produce the Thz current by beating with the $a_2^*$ quiver, 
 $B_2 =C (ck_2 + ck_3)(ck_1+ck_2)/(\omega_1-\omega_3)^2$ represents the ponderomotive force by  $a_3 a_2^* $,producing  the THz current by beating with the $a_1$ quiver and  $n_L$.% which will produce the Thz current by beating with the $a_1$ quiver, 
$B_3 =C (ck_1 - ck_3)(ck_1+ck_2)/(\omega_2-\omega_3)^2$ represents  the ponderomotive force by $a_3 a_1^* $, producing the THz current by beating with the $a_2$ quiver and  $n_L^*$ and 
%which will produce the Thz current by beating with the $a_2$ quiver, 
 $B_4 =C (ck_2 - ck_3)(ck_1+ck_2)/(\omega_1+\omega_3)^2$ represents the ponderomotive force by  $a_3 a_2 $,  producing the THz current by beating with the $a_1^*$ quiver and   $n_L^*$.
% which will produce the Thz current by beating with the $a_1^*$ quiver. 
Combining Eq.~(\ref{eq:first}) and Eq.~(\ref{eq:second}), we obtain 
the total density response if we replace 
  $A_{1,2,3,4}$ by $ D_{1,2,3,4} = A_{1,2,3,4} - B_{1,2,3,4}$. 
Combining  Eq.~(\ref{eq:first}) and Eq.~(\ref{eq:second}) with Eq.~(\ref{eq:1}) or Eq.~(\ref{eq:2}), 
we obtains  

\begin{eqnarray}
 L_3 a_3  =  \frac{\omega_{\mathrm{pe}}^2 }{2\omega_3} \left( 
\Gamma C_B |a_1|^2 |a_2|^2 \right)a_3  \nonumber \\ \nonumber \\
=  \frac{\omega_{\mathrm{pe}}^2 }{2\omega_3} \frac{n_L}{n_0}\left( 
\Gamma  |a_1| |a_2| \right)a_3 
\mathrm{,}\label{eq:major}  \\ \nonumber 
\end{eqnarray}
where $ \Gamma = D_3 + D_4 - D_1 - D_2$ and $|n_L| \cong C_B |a_1^*a_2|$. 
%If $\omega_1 > \omega_2$, it can be shown that the THz will decay. 
The equation~\ref{eq:major} is the major result of this paper. 
As an example, consider $\omega_1 = 5\omega_3 $ and $\omega_2= 6 \omega_3$. 
Then, 
$A_1 = 0.73,  A_2 =3.06,  A_3 =0.64, A_4=0.69$,
$B_1 = 1.35,  B_2 =4.81,  B_3 =1.76, B_4=1.53$
%$A_1 = 1.06,  A_2 =2.13,  A_3 =0.92, A_4=0.48$,
%$B_1 = 0.32,  B_2 =0.67,  B_3 =0.42, B_4=0.21$
and $\Gamma = 0.4$, which is positive. 
There will be an amplification.
As an another example, consider $\omega_1 = 10 \ \omega_3 $ and $\omega_2= 11 \ \omega_3$. 
Then, 
%$A_1 = 1.03,  A_2 =1.45,  A_3 =0.96, A_4=0.69$,
%$B_1 = 0.17,  B_2 =0.25,  B_3 =0.20, B_4=0.14$ and 
$\Gamma =  0.1$. 
As an another example,  consider $\omega_1 = 15 \ \omega_3 $ and $\omega_2= 16 \ \omega_3$. 
%Then, $A_1 = 1.02,  A_2 =1.28,  A_3 =0.98, A_4=0.78$,
%$B_1 = 0.12,  B_2 =0.15,  B_3 =0.13, B_4=0.10$
%and 
$\Gamma = 0.05$. 
The more frequency difference between the THz  light and the lasers, the efficiency for  the amplification becomes weaker. 
If $\omega_1 < \omega_2 $, then  $\Gamma $ will be negative and there will be decay. 
It can be concluded that the excitation of the Langmuir wave by the counter-propagating lasers could amplify the THz light if the THz is co-propagate with the lower frequency laser.

For the example with the specific parameter, consider a plasma with $n_0 = 10^{17}  \ / /\mathrm{cm}^3$,  $n_L / n_0 =0.3$, a  6 THz light and two lasers with $10 \ \mu \mathrm{m} $ and  $9.0 \ \mu \mathrm{m} $. 
Then, it can be estimated as $\Gamma  \cong 1.5 \times 10^{11} \sqrt{I^1_{16}I^2_{16} } \sec$, where $I^1_{16}$ ($I^2_{16}$) is the intensity of $a_1$ ($a_2$) laser normalized by $10^{16} \ W /\mathrm{cm}^2$. For the relativistic intensity, the gain-per-length could reach 10 per centimeter. 
As an another example, consider a plasma with $n_0 = 4 \times 10^{17}  \ / /\mathrm{cm}^3$,  $n_L / n_0 =0.3$, a  10 THz light and two lasers with $10 \ \mu \mathrm{m} $ and  $8.5 \ \mu \mathrm{m} $. 
Then, it can be estimated as $\Gamma  \cong 3 \times 10^{12} \sqrt{I^1_{16}I^2_{16} } \sec$.
 For the relativistic intensity, the gain-per-length could reach 100 per centimeter. 

To summarize, 
in this paper, a new amplifying mechanism of the THz light is proposed based on the Raman scattering. The desirable phase-locking is provided by the pre-existing Langmuir wave excited by the BRS.  As the plasma density responds to the ponderomotive interaction between the THz light and the lasers under the pre-existing Langmuir wave,   the lasers could give away energy into the THz light via the non-resonant Raman scattering.  The THz light needs to be propagating with the low-frequency laser. The gain per length could reach 100 per centimeter. 

If the THz light is propagating with the low-frequency laser, it is also beneficial. 
As the Langmuir wave is excited by the BRS between lasers,  the energy will be also transferred from the higher frequency laser to the lower one. 
The lower frequency laser will be amplified as it extract the energy from the fresh higher frequency laser because the lasers are counter-propagating. The THz light will also benefit from this amplified lower frequency laser because it is co-propagating  and experiencing more intense laser.

\bibliography{tera2}% Produces the bibliography via BibTeX.

\end{document}